\documentclass[]{aastex631}
\usepackage{amsmath}
\submitjournal{ApJ}

\newcommand{\Mbh}{M_\bullet}

\begin{document}

\title{Energy Flux and Particle Flux in Steady-State Solutions of Nuclear Star Clusters}

\correspondingauthor{Barak Rom}
\email{barak.rom@mail.huji.ac.il}

\author[0000-0002-7420-3578]{Barak Rom}
\affiliation{Racah Institute of Physics, The Hebrew University of Jerusalem, 9190401, Israel}
\author[0000-0002-8304-1988]{Itai Linial}
\affiliation{Institute for Advanced Study, 1 Einstein Drive, Princeton, NJ 08540, USA}
\affiliation{Department of Astronomy, Columbia University, New York, NY 10027, USA}
\author[0000-0002-1084-3656]{Re'em Sari}
\affiliation{Racah Institute of Physics, The Hebrew University of Jerusalem, 9190401, Israel}

\shorttitle{Steady-state fluxes in nuclear star clusters}
\shortauthors{Rom, Linial \& Sari}

\begin{abstract}
\noindent We examine the effects of two-body interactions in a nuclear star cluster surrounding a supermassive black hole. We evaluate the energy flux, analogously to the particle flux calculation of \cite{BW_76}. We show that there are two types of power-law steady-state solutions: one with zero energy flux and constant particle flux and the other with constant energy flux and zero particle flux.
We therefore prove that a zero particle flux solution, which corresponds to the case of an accreting supermassive black hole, can be obtained by requiring a constant energy flux. Consequently, this solution can be derived by simple dimensional analysis, bypassing the need for detailed calculation.
Finally, we show that this characteristic, of zero particle flux for constant energy flux and vice versa, is not unique to the Keplerian potential of a supermassive black hole but holds for any central potential of the form $\phi\propto r^{-\beta}$.
\end{abstract}


\section{Introduction} \label{sec:intro}
Supermassive black holes (SMBHs) in the center of galaxies are surrounded by nuclear star clusters that consist of millions to billions of stars and an unknown number of stellar-mass to intermediate-mass black holes. 
This dense environment nourishes a variety of astrophysical phenomena, such as tidal disruption events \citep[TDEs;][]{Rees_88,TDE_rev,Velzen_2021}, quasiperiodic eruptions \citep[QPEs;][]{Miniutti_2019,Arcodia_2021}, and extreme-mass-ratio inspirals \citep[EMRIs;][]{Babak_17}. Therefore, there is a special interest in the study of the dynamics in such clusters. Specifically, the steady-state distribution of the orbital elements is important to understand the above physical phenomena.


Generally, a steady-state solution implies constant particle and energy fluxes. \cite{Peebles_72} showed that assuming spherical symmetry and a density distribution of the form $n(r)\propto r^{-\alpha}$, the steady-state solution for a single mass population is obtained by requiring $\alpha=9/4$. This can be understood by the following simplified derivation; the ``na\"ive" particle flux is estimated as the total number of particles, $N(r)$, times the scattering rate, $\Gamma(r)\sim n(r)\sigma v$, 
\begin{equation} \label{eq:int1}
\mathcal{F}_p\sim N(r)\Gamma(r)\propto \frac{n^2(r)r^3}{v^3}.
\end{equation}
Requiring $\mathcal{F}_p={\rm const.}$ leads to 
\begin{equation}\label{eq:int1a}
n(r)\propto r^{-9/4},
\end{equation}
where we consider particles well within the radius of influence and therefore the gravitational potential is dominated by the SMBH.

\cite{BW_76} revisited \cite{Peebles_72} argument and, by a precise calculation of the net scattering rate, concluded that the steady-state density distribution leads to $\alpha=7/4$. Their detailed calculation showed that this profile satisfies an exact balance between the inward and outward scatterings and so the particle flux vanishes. Shallower profiles lead to an inward flux of mass, while steeper slopes result in an outward flux. 
A heuristic argument \citep{Shapiro_76,Biney_Tremaine,Sari_Gold_06,AH_09,Sari_Fragione_2018} suggests that this profile can be derived more easily by considering the energy flux. As the energy per star is much larger near the SMBH than in the outskirts of the cluster, a consistent solution can be obtained by demanding a constant energy flux and a vanishing particle flux, i.e., much smaller than the na\"ive estimation given in Eq. (\ref{eq:int1}). The energy flux can be evaluated as
\begin{equation} \label{eq:int2}
\mathcal{F}_E\sim N(r)E(r)\Gamma(r)\propto \frac{n^2(r)r^2}{v^3}.
\end{equation}
Requiring that $\mathcal{F}_E={\rm const.}$ leads to the known result
\begin{equation} \label{eq:int2a}
n(r)\propto r^{-7/4}.
\end{equation}
Thus, the simple constraint on the energy flux leads to the \cite{BW_76} profile.

The steady-state calculation was generalized for multiple mass component populations \citep{BW_77,AH_09,Keshet_2009,Vasiliev_17,Sari_Linial_22} and has been widely used to evaluate the rates of TDEs and EMRIs \citep{Preto_Amaro_Seoane_2010,Merritt_2010,Merritt_etal_2010,Aharon_Perets_2016,Panamarev_2019,Stone_2020,Emami_2021,BroBorBon22}.

As evident from Eqs. (\ref{eq:int1}) and (\ref{eq:int2}), given a power-law distribution the particle flux and the energy flux scale differently, namely, only one of the fluxes can be a nonzero constant. Therefore, in a steady-state solution, the other flux must vanish. This sort of zero-flux solution is possible since in the above flux calculations there is a missing coefficient of order unity that cannot be determined by order of magnitude estimates yet may vanish for specific power-law indices of the distribution, as happens for the particle flux in the \cite{BW_76} profile.

Consequently, we distinguish between two types of steady-state solutions, type $I$, in which the particle flux is zero and there is a constant energy flux, as in the \citet{BW_76} profile, and type $II$, where the energy flux is zero and there is a constant particle flux, as \citet{Peebles_72} type profiles.

In this paper, we prove by explicit calculation that when the energy flux is constant the particle flux vanishes and vice versa. In section \ref{sec:en_flux} we calculate the energy flux by integrating over all two-body scatterings, analogously to the particle flux calculation of \citet{BW_76}. In section \ref{sec:gp} we generalize the calculation to a central potential of the form $\phi\propto r^{-\beta}$, with $\beta\neq0$, and show that the two types of steady-state solutions are not a unique feature of the Keplerian potential of an SMBH but a general characteristic of the scale-free dynamics.

\section{Energy Flux} \label{sec:en_flux}
We consider an SMBH of mass $\Mbh$, surrounded by a population of stars with mass $m$ and specific energy $E_i=G\Mbh/r_i-v^2_i/2$. As \citet{BW_76}, we assume (a) a spherically symmetric spatial distribution, (b) isotropic velocities, and (c) that the dynamics are dominated by weak scatterings.
The particle and energy fluxes are calculated by integrating over all Newtonian interactions between two particles - with arbitrary initial and final energies, $\left(E_a, E_b\right)\rightarrow\left(E'_a, E'_b\right)$ - while taking into account the interaction rate and the conservation of energy and momentum. The interaction rate per star is $n\sigma v_r$, where $n$ is the number density of the scatterers, $v_r$ is the relative velocity, and $\sigma$ is the gravitational Rutherford cross section \citep{LL_Mech}, which for small scattering angles is given by $\frac{d\sigma}{d\Omega}\approx\frac{16G^2m^2}{v_r^4\theta^4}$, where $\theta$ is the scattering angle.
Finally, for the particle flux through a given energy $E$, only interactions where one of the particles crosses $E$ are counted. After integration over all possible interaction orientations\footnote{For further details see section III.A in \cite{BW_76}.}, the particle flux can be expressed as a simple integration over the energy, as given in Eq. $(40)$ of \citet{BW_76}\footnote{In the \cite{BW_76} paper, $\mathcal{F}_{P}$ is denoted as $R(E)$.}
\begin{equation} \label{eq:en0}
    \mathcal{F}_P=\frac{32\sqrt{2}\pi^5}{3}G^5m^2\Mbh^3\ln\left(\Lambda\right)\int_0^\infty dE_b\max\left\{E,E_b\right\}^{-3/2}\Big[f\left(E\right)f'\left(E_b\right)-f\left(E_b\right)f'\left(E\right)\Big],
\end{equation}
where $\ln\left(\Lambda\right)$ is the Coulomb logarithm, with $\Lambda\sim\Mbh/m$ (typically for the center of galaxies $\ln\left(\Lambda\right)\approx10$). The distribution function in phase space \citep{Biney_Tremaine}, $f\left(E\right)$, is defined such that $n(r)=\int d^3v f\left(E\right)$.

Assuming that the distribution function has a power-law form, $f(E)=f_0\left(E/E_0\right)^p$, we can integrate Eq. (\ref{eq:en0}) to obtain
\begin{equation} \label{eq:en01}
    \begin{gathered}
    \mathcal{F}_P=DE^{2p-3/2}\frac{3\left(1/4-p\right)}{p\left(p+1\right)\left(p-1/2\right)\left(p-3/2\right)},\\
    D\equiv\frac{32\sqrt{2}\pi^5}{3}\frac{pf_0^2}{E_0^{2p}}G^5m^2\Mbh^3\ln\left(\Lambda\right),
    \end{gathered}    
\end{equation}

We calculate the energy flux by decomposing it into two parts. The first part is the advective term, corresponding to the energy carried by particles that cross a given energy level and is, therefore, proportional to the particle flux. The second part is the conductive term, accounting for the energy transferred by interactions between particles above the given energy and particles below it. A similar distinction can be found in \cite{Vasiliev_17}, where the derivation is performed in an equivalent Fokker-Planck formalism.

Using Eq. (\ref{eq:en01}), the advective term can be easily determined:
\begin{equation} \label{eq:en1}
    \begin{gathered}
   \mathcal{F}_{E,adv}=E\cdot \mathcal{F}_P=DE^{2p-1/2}\frac{3\left(1/4-p\right)}{p\left(p+1\right)\left(p-1/2\right)\left(p-3/2\right)},\\        
    \end{gathered}
\end{equation}

The calculation of the conductive term retraces \citet{BW_76} calculation of the particle flux, as described above. It is illuminating to compare the power in which the energy difference, $\Delta\equiv E'_a-E_a$, appears in the particle flux and the energy flux calculations:
(a) While only interactions where a particle crosses the given energy level $E$ contribute to the particle flux, all of the interactions between particles above the given energy level with particles below it contribute to the conductive energy flux. Therefore, for a given transferred energy $\Delta$, the conductive energy flux takes into account $\sim E/\Delta$ more encounters compared to the particle flux.
(b) The calculation of the energy flux includes an extra power of $\Delta$, compared to the particle flux, as we consider the energy transfer in each interaction.
The combination of the points (a) and (b) gives an extra power of $E$ to the energy flux compared to the particle flux, as expected from Eqs. (\ref{eq:int1}) and (\ref{eq:int2}), while keeping the same factor of $1/\Delta$ in both of the fluxes, leading to the same logarithmic divergence.

Thus, the conductive energy flux is determined by integrating over all relevant interactions, given the interaction rate per star and the abovementioned assumptions, and it is given by
\begin{equation} \label{eq:en2}
\mathcal{F}_{E,cond}=-\frac{32\sqrt{2}\pi^5}{3}G^5m^2\Mbh^3\int_{\Delta_{min}}^{\Delta_{max}} \frac{d\Delta}{\Delta}\int_E^\infty \frac{dE_b}{E_b^{3/2}}\int_0^E dE_a\Big[f(E_a)f'(E_b)-f(E_b)f'(E_a)\Big],
\end{equation}
where, without loss of generality, we assume $E_a<E$ and $E_b>E$. Note that this choice, together with the definition of $\Delta$, introduces the minus sign in Eq. (\ref{eq:en2}).


Assuming that $f(E)$ has a power-law form, the integration over the energies can be done, in addition to the trivial integration over $\Delta$, and we obtain
\begin{equation} \label{eq:en3}
\mathcal{F}_{E,cond}
=DE^{2p-1/2}\frac{3/2}{p\left(p+1\right)\left(p-1/2\right)\left(p-3/2\right)}.
\end{equation}
Adding Eqs. (\ref{eq:en1}) and (\ref{eq:en3}) we get that the total energy flux is
%
\begin{equation} \label{eq:en4}
\mathcal{F}_E=\mathcal{F}_{E,adv}+\mathcal{F}_{E,cond}=DE^{2p-1/2}\frac{3\left(3/4-p\right)}{p\left(p+1\right)\left(p-3/2\right)\left(p-1/2\right)}.
\end{equation}

Note that the particle flux and the energy flux, Eqs. (\ref{eq:en01}) and (\ref{eq:en4}), respectively, converge for $p<1/2$. Moreover, it is evident that
\begin{equation} \label{eq:en4a}
    \begin{gathered}
    \mathcal{F}_P\propto E^{2\left(p-3/4\right)}\left(1/4-p\right), \\        
    \mathcal{F}_E\propto E^{2\left(p-1/4\right)}\left(3/4-p\right),
    \end{gathered}
\end{equation}
and therefore $p=1/4$, which is the \citet{BW_76} solution, which satisfies a constant energy flux and a zero particle flux. The value of the constant flux can be easily calculated using Eq. (\ref{eq:en4})
\begin{equation} \label{eq:en5}
\mathcal{F}_E=\frac{1024\sqrt{2}}{25}\left(\frac{f_0}{E_0^{1/4}}\right)^2\pi^5G^5m^2\Mbh^3\ln\left(\Lambda\right),
\end{equation}
where $f_0$ and $E_0$ are determined by the normalization of the distribution function. This calculation, just like that of \cite{BW_76} is accurate up to a precision of $1/\ln\left(\Lambda\right)\sim10\%$, due to the contribution of strong interactions. This result agrees with the result of \cite{Vasiliev_17}.

The second possibility of a nonzero constant particle flux and zero energy flux requires $p=3/4$, as \cite{Peebles_72} showed. However, under the assumption of isotropic velocity distribution, this solution leads to diverging particle and energy fluxes, and therefore its physical meaning is unclear.
Yet, in the next section, we generalize our calculation to other central potentials where these divergences are eliminated.
\section{General Central Potential} \label{sec:gp}
In order to show that the two types of steady-state solutions are a general characteristic of the dynamics, we consider a central potential of the form $\phi\propto r^{-\beta}$, with $\beta\neq0$.
Examples of systems with non-Keplerian potential are the cores of globular clusters and the outskirts of nuclear star clusters, i.e., beyond the SMBH's radius of influence \citep{Shapiro_76,Shapiro_78}. Although in these examples the potential does not have a power-law form\footnote{In the mentioned cases, the distribution function can be taken as isothermal, $f(E)\propto e^{-\beta E}$, and therefore the energy flux and the particle flux both vanish, as evident from Eqs. (\ref{eq:en0}) and (\ref{eq:en2}).}, we consider the above simple potential to verify the generality of the zero particle flux and zero energy flux dichotomy.
Given this new potential, the na\"ive fluxes, analogously to Eq. (\ref{eq:int1}) and (\ref{eq:int2}), are proportional to
\begin{equation}\label{eq:gp1}
\begin{gathered}
    \mathcal{F}_P\propto r^{-2\alpha+(6+3\beta)/2}, \\
    \mathcal{F}_E\propto r^{-2\alpha+\left(6+\beta\right)/2}.
\end{gathered}
\end{equation}

Retracing the quantitative calculation presented in section \ref{sec:en_flux}, with the new central potential, $\phi\propto r^{-\beta}$, we obtain that the particle flux is given by
%
\begin{equation}\label{eq:gp3}
\mathcal{F}_P\propto E^{2p-3/\beta+3/2}\frac{3\left(1-2/\beta\right)\big[p+5/4-3/\left(2\beta\right)\big]}{p\left(p+1\right)\left(p+3/2-3/\beta\right)\left(p+5/2-3/\beta\right)}.  
\end{equation}

There are two possible solutions for a constant particle flux:
\begin{equation}\label{eq:gp4}
\begin{aligned}
    p_1&=\frac{3}{2\beta}-\frac{5}{4}\quad\rightarrow\quad\alpha_{1}=\frac{6+\beta}{4},\\
    p_2&=\frac{3}{2\beta}-\frac{3}{4}\quad\rightarrow\quad\alpha_{2}=\frac{6+3\beta}{4},
\end{aligned}
\end{equation}
where $p$, the exponent of $f(E)$, corresponds to $\alpha$, the exponent of $n(r)$, according to the relation $n(r)\propto f(E)v^3$. 

The exponents $p_1$ and $p_2$ are the generalized forms of the \citet{BW_76} and \citet{Peebles_72} solutions, respectively. The first solution, $p_1$, corresponds to a zero particle flux since the term in the square brackets in Eq. (\ref{eq:gp3}) vanishes. The second solution, $p_2$, gives a nonzero constant particle flux, as it nulls the exponent of the scaling relation in Eq. (\ref{eq:gp3}). 

The modified energy flux is given by
\begin{equation}\label{eq:gp5}
\mathcal{F}_E\propto 
E^{2p-3/\beta+5/2}\frac{3\left(1-2/\beta\right)\big[p+3/4-3/\left(2\beta\right)\big]}{p\left(p+1\right)\left(p+3/2-3/\beta\right)\left(p+5/2-3/\beta\right)}.
\end{equation}

Thus, the particle flux and the energy flux can be written as follows:
\begin{equation} \label{eq:gp6}
    \begin{gathered}
    \mathcal{F}_P\propto E^{2\left(p-3/\left(2\beta\right)+3/4\right)}\left(p-\frac{3}{2\beta}+\frac{5}{4}\right), \\        
    \mathcal{F}_E\propto E^{2\left(p-3/\left(2\beta\right)+5/4\right)}\left(p-\frac{3}{2\beta}+\frac{3}{4}\right).
    \end{gathered}
\end{equation}
Note that we assume $p+5/2-3/\beta<0$ and therefore both of the fluxes converge.
From Eq. (\ref{eq:gp6}) it is clear that $p_1$, as defined in Eq. (\ref{eq:gp4}), leads to a nonzero constant energy flux and a vanishing particle flux, while $p_2$ leads to the opposite scenario, a vanishing energy flux and a constant particle flux.

\section{Summary \& Discussion} \label{sec:con}
We show by a detailed calculation of the energy flux, combined with \cite{BW_76} calculation of the particle flux, that there are two types of steady-state solutions: one with zero energy flux and constant particle flux, and the other with constant energy flux and zero particle flux. 

Our calculation validates the derivation of the zero particle flux steady-state distribution, which corresponds to the case of accretion to the SMBH, from the dimensional analysis of the energy flux 
\citep[as suggested in][]{Shapiro_76,Biney_Tremaine,Sari_Gold_06,AH_09,Sari_Fragione_2018}.
Similarly, it shows that the zero energy flux solution could be obtained from dimensional analysis of the particle flux. This solution is relevant for decretion particle flow \citep[see, for example,][]{Sari_Fragione_2018}.
We strengthen this understanding by showing that this is a general characteristic of the two-body dynamics in a central potential and not a unique feature of the Keplerian potential of an SMBH.

Finally, as shown by \citet{BW_76}, the flux calculations, under the abovementioned assumptions, lead to unphysical divergences in steep density profiles, with $p>1/2$, such as the \cite{Peebles_72} solution. Note that although it seems from Eq. (\ref{eq:en4}) that the energy flux vanishes for $p=3/4$, it actually diverges since the integrals in Eqs. (\ref{eq:en1}) and (\ref{eq:en3}) do not converge. 
This divergence stems from the assumption of isotropic velocity distribution, also known as the thermal eccentricity distribution, which leads to the presence of particles in highly eccentric orbits that can interact with energetic particles near the SMBH. In order to remedy this unphysical behavior, one should consider the angular momentum diffusion and solve the full $2d$ diffusion equations in the energy and angular momentum space, which is beyond the scope of this work. However, for weaker central potential, with $0<\beta<6/7$, both the constant energy flux and constant particle flux solutions are consistent with the assumption of isotropic velocity distribution, as evident from Eqs. (\ref{eq:gp3}) and (\ref{eq:gp5}).

\begin{acknowledgments}
This research was partially supported by an ISF grant, an NSF/BSF grant, and an MOS grant. B.R. acknowledges support from the Milner Foundation. I.L. thanks support from the Rothschild Fellowship and the Gruber Foundation.
\end{acknowledgments}


\bibliography{main}{}
\bibliographystyle{aasjournal}

\end{document}